# COMPUTER AIDED TOLERANCING BASED ON ANALYSIS AND SYNTHETIZES OF TOLERANCES METHOD.

A. Hassani N. Aifaoui & A. BenAmara
*LGM, ENIM, 05 Avenue Ibn El Jazzar – Monastir 5019*
*h_abdou4@yahoo.fr; Nizar.Aifaoui@ipeim.rnu.tn, abdel.benamara@enim.rnu.tn*

S. Samper
SYMME, Polytech'Savoie, 74940 Annecy-le-Vieux
*Serge.Samper@univ-savoie.fr*

**ABSTRACT**

The tolerancing step has a great importance in the design process. It characterises the relationship between the different sectors of the product life cycle: Design, Manufacturing and Control. We can distinguish several methods to assist the tolerancing process in the design. Based on arithmetic and statistical method, this paper presents a new approach of analysis and verification of tolerances. The chosen approach is based on the Worst Case Method as an arithmetic method and Monte Carlo method as a statistical method. In this paper, we compare these methods and we present our main approach, which is validated using an example of 1 D tolerancing.

## I. INTRODUCTION

The Product Life Cycle is an association of many stages. Generally, we find three steps: Design, Manufacturing and Control. The concept of tolerancing is the link vector between these steps. The following figure introduces the mechanical product life cycle and presents tolerancing as a shared data between several actors.

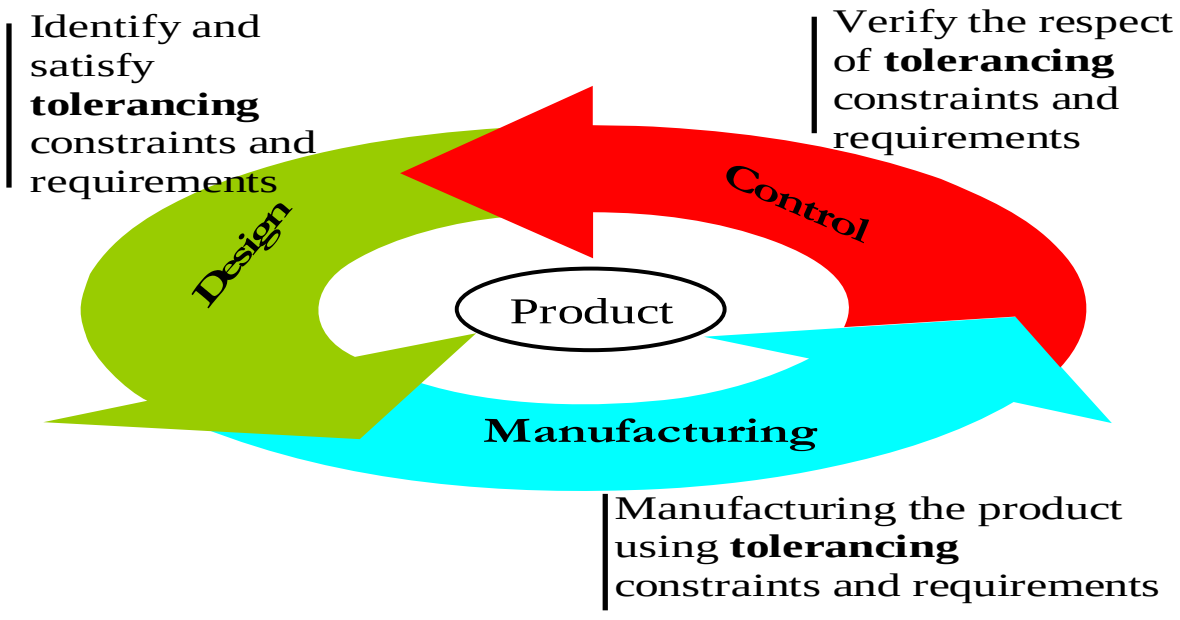

**Figure 1. Different steps of product life cycle.**

A first step of product creation is beginning with idea. The achievement of this idea is limited by many constraints in different steps. Under these constraints and conditions, we find thus imposed by the customer, thus imposed by the manufacturing process, …[1, 2]. These constraints are represented by the dimension limits or variation ratings. This representation is clear, easy and comprehensive by different stakeholders. It characterizes the state of surfaces of a product and their positioning in the form of dimensional and geometric constraints. This state defines the domain of tolerancing. This domain is the subject of several studies and researches based on various methods. The main objective of these studies is ensuring customer satisfaction while ensuring a better quality and low cost. To ensure an optimal inclusive cost/quality, the tolerancing is to study the variation of dimensional and geometrical limits of products. A production is the result of a well studied approach that requires knowledge of the limits of variation of each side to ensure the proper functioning of the mechanism and give a margin of adapting production to the requirements and constraints. This state needs to be allocated to each gap differences and avoid constraints and quality. To respond to these points, several methods are involved whatsoever arithmetic methods or statistical methods [3, 4, 5]. Among the methods used, we limit ourselves to two methods which are the Worst Case method as an arithmetic method and the Monte Carlo method as a statistical method [2,3,4,5,6,7]. What allows identifying the characteristics and main features of each method while providing a comparison of the usefulness and validity of each.

## II. Background

The study of the variation in size and geometry of dimensions in a product is relying on a study of the chain of ratings. It is the overlapping parts one over the other to ensure the functioning of a system. The following figure shows an example of chain dimensions;

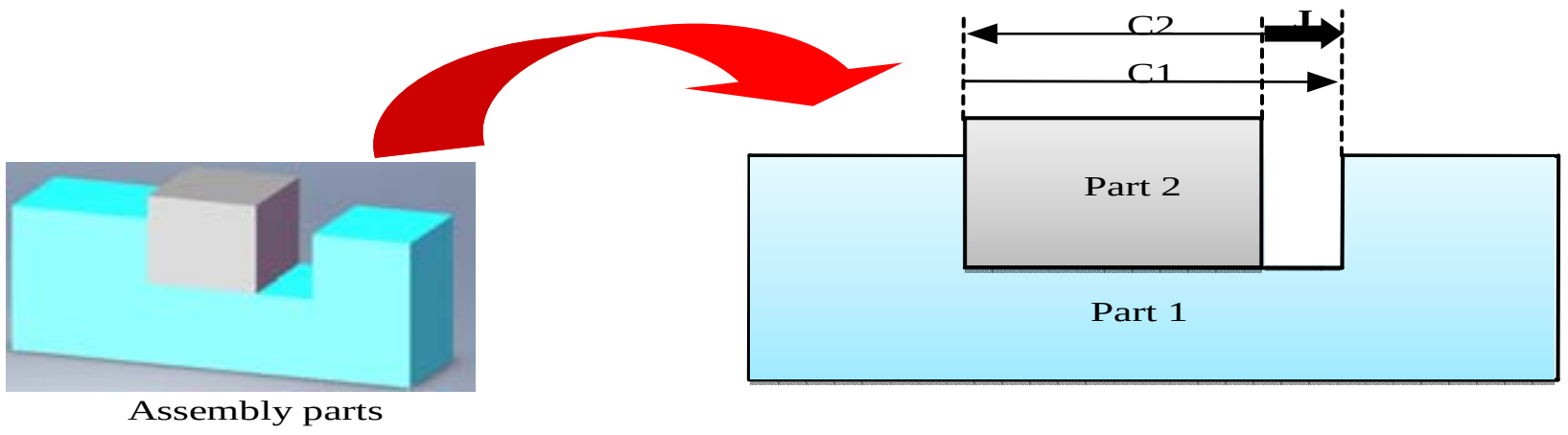

**Figure 2. Ratings Chain of an assembly parts**

This example presents a ratings chain an assembly including two parts (Part 1 and Part 2) providing a set of guiding 'J' in translation. This functional condition is satisfied using the following relation:

$$J = C1 - C2$$

## III. APPLIED METHODS

### *a- Worst Case Method (WCM)*

It is a method to determine and calculate the Functional Condition (FC) and other rating in their limits and this variation from different components, participants in a chain of ratings. This chain is based on the following relationships (J = C1 – C2) to identify the minimum and maximum values of each element. This method is used to define extreme values ratings without giving details on what happens inside tolerances zone of each rating.
The conformity of this method is defined by an interval (min, max), which exceeds its share or other area is the rejection or non-compliance of parts. Moreover, this method does not provide information on combinations of scores to secure the assembly.
So to address these problems, one is led to research other methods whose objectives are:

- Know what is inside the tolerances interval.
- Try to extend the zone of acceptance in order to reduce constraints on.
- Present the various possible combinations between the players ratings product

The cure of disadvantages of this method is modelled in statistical method. This method is the Monte Carlo method.

### *b- Monte Carlo Method (MCM)*

This method is a statistical method which requires its application;

- To characterize and identify the problem of study followed by sampling, i.e through iterations on achievements.
- Operating the result.

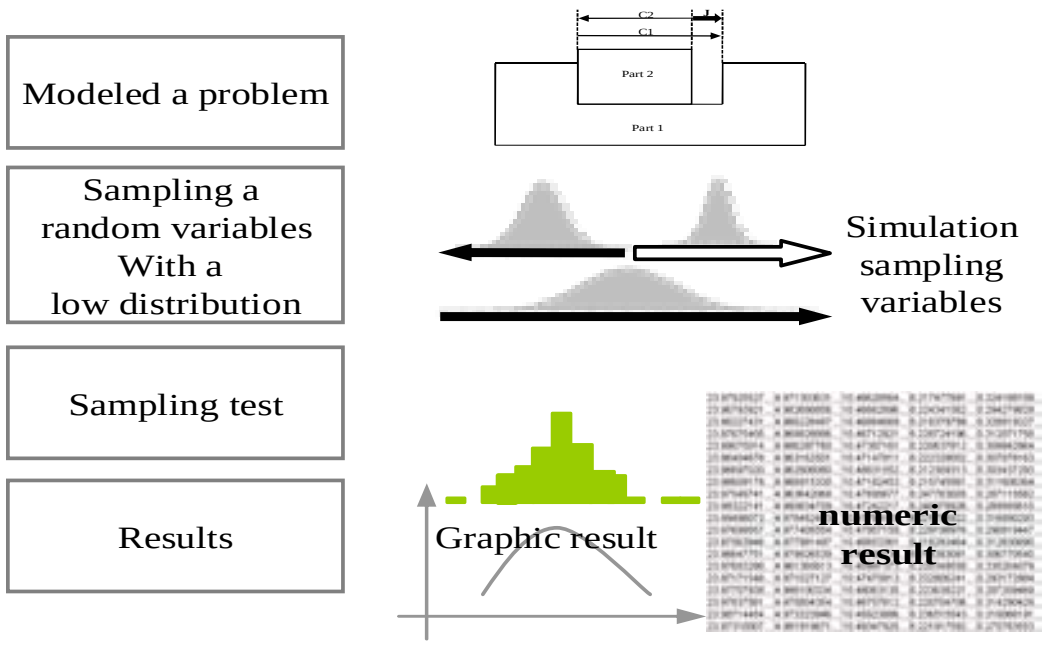

**Figure 3. Monte Carlo algorithm**

Apply these two methods to bring and simulate manufacturing from the moment of conception whose objectives are;

- Choose the one hand tolerances ratings and other identifying values ratings in a tolerancing assistance and aid to the choice of tolerances based on two methods whether the WCM or MCM.
- Getting a comprehensive approach to specification and verification of tolerances.

The application of this method is based the next algorithm:
For these two methods, this paper presents an algorithm describing their applications in the field of specification and verification of tolerances.

***c- Assumptions***

The disposal can be changed according to the requirements of customers.
Thus the application of this approach is based on the following assumptions;

- To consider the 1-dimensional chain
- The chain of ratings is linear and dimensional
- The principle of independence will be preserved
- The pieces studied are considered solid and rigid parts. So the deformability of parts is not considered in this step of study.

The main goal in this stage is to apply these methods to analyse and verify the tolerance of parts in assemblies. Based on assumptions, this paper verifies the compliance of assembly and the guarantee of constraints and requirements. In the following paragraph, the algorithm of applied method is presented.

## IV. PROPOSED METHOD

This paper presents a new approach to analyse and verify tolerances in their domain of variation. This study gets an assistant aided by Designer to allocate a tolerance of parts in assembly. The main goal of this step is to reduce a cost of product. In the first part, the identification of a value of rating or a value of Functional rating depends on the same ratings of chain. Traditionally, this value is defined by arithmetic method with Worst Case method. Recently, some statistical methods are applied to solve a problem of tolerancing [3, 4, 5, 6, 7]. Several CAT tools are created; NX Quickstack, ToleranceCal, Mecamaster, CATIA Tolerancing, TOLTECH, Ce/Tol 6 Sigma, Tolerance Manager de PCO, eM-Tolmate, etc. [8].
In this paper, we propose a solution to study a tolerancing domain. In the first part, we analyzed a chain of tolerance to search a Functional rating or other rating in chain. In the second part, we verified a compliance of chain respecting an imposed rating or a Functional rating.
For the analysis of tolerances, the data will be the nominal value and deviations ratings of the chain. The result is the extreme value of the ratings (figure 4.A).

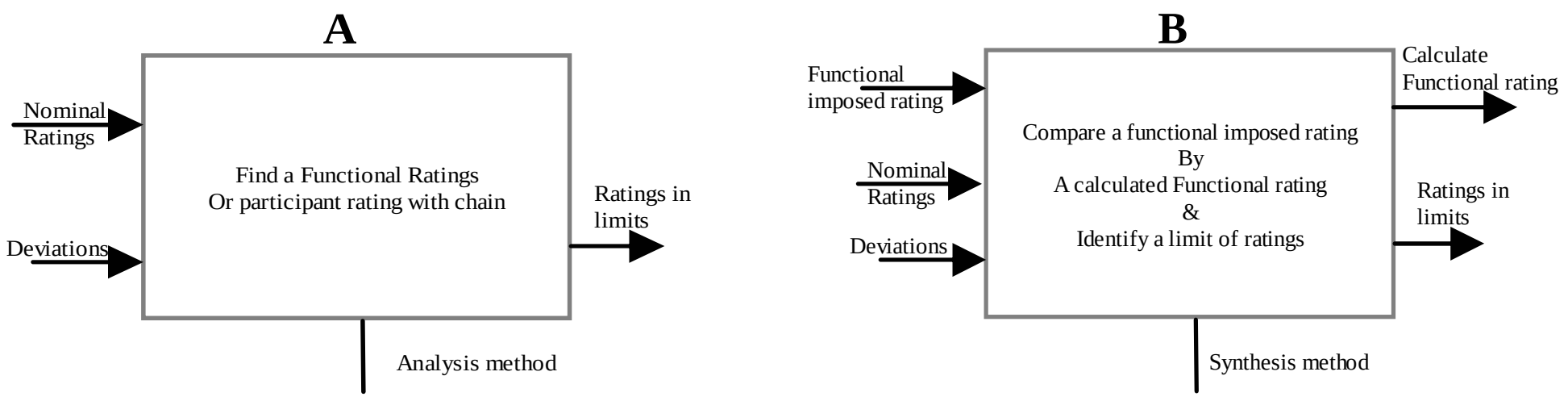

**Figure 4. Tolerancing method**

For the synthesis phase (figure 4-B), a functional condition will be imposed in order to be respected while introducing the values of ratings and their deviations to ensure the guarantee of assemblies and ensure compliance of the condition imposed. This method defines the compliance by an interval. The compliance is limited by the extreme values of Functional imposed rating (figure 5). Outside this interval is the domain of non compliance. So for two methods, it begins by introducing the nominal values of dimensions and differences in the analysis. And for the synthesis, we introduce the requirement to respect the functional uses and one of two methods;

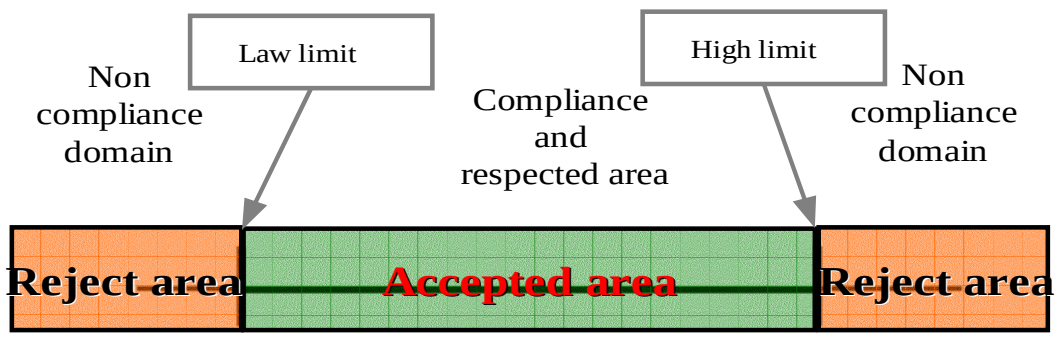

**Figure 5. Compliance domain**

### *a- Worst case algorithm*

The calculation by the WCM can determine and / or verify their extreme values and their compliance. The algorithm (figure 6) describes the step to determine the extreme values of different ratings in the chain of ratings. It presents also the step of verification of the Functional condition (rating). The compliance of rating is defined when the calculated rating is included in the limit of imposed Functional condition. This method can not determine different combinations between the participant elements in the chain. A new research deals with tolerancing relying on statistical methods [3, 4, 5, 6, 7, 9, 10, 11]. Among the methods cited in literature, this paper presents a solution of tolerancing problem by Monte Carlo method. The next paragraph presents the application of Monte Carlo method. The main objective of this method is;

- Giving an idea about the actor of product on the various combinations between the different constituents of ratings of a chain or assembly.
- Extend the zone of acceptance.
- Giving an idea to know what is happening inside each interval.

### *b- Monte Carlo algorithm*

The MCM is based on sampling according to a well-defined law. This paper presents an algorithm of application of Monte Carlo method. This method can be sacrificed to a low probability of defective parts in the manufacturing assembly. In this paper, we limited our study to the normal distribution as a law of iteration [11, 12]. This normal distribution is characterized by a mean and standard deviation $(\sigma)$ or variability $(\sigma^2)$. Based on the assumptions, the parameters of sampling are independent variables. In this stage, the mean is; $\mu = \sum \alpha_i \cdot \mu_i$ and the standard deviation is; $\sigma = \sqrt{\sum \alpha_i^2 \cdot \sigma_i^2}$ . These parameters characterized a normal distribution in the assembly.

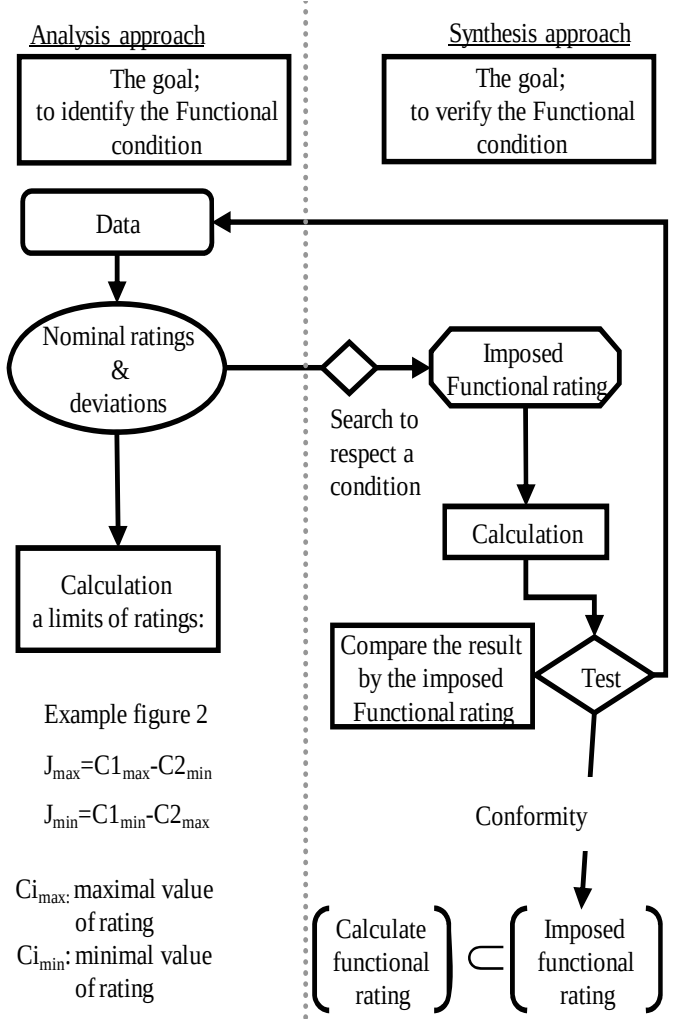

**Figure 6. Worst Case algorithm**

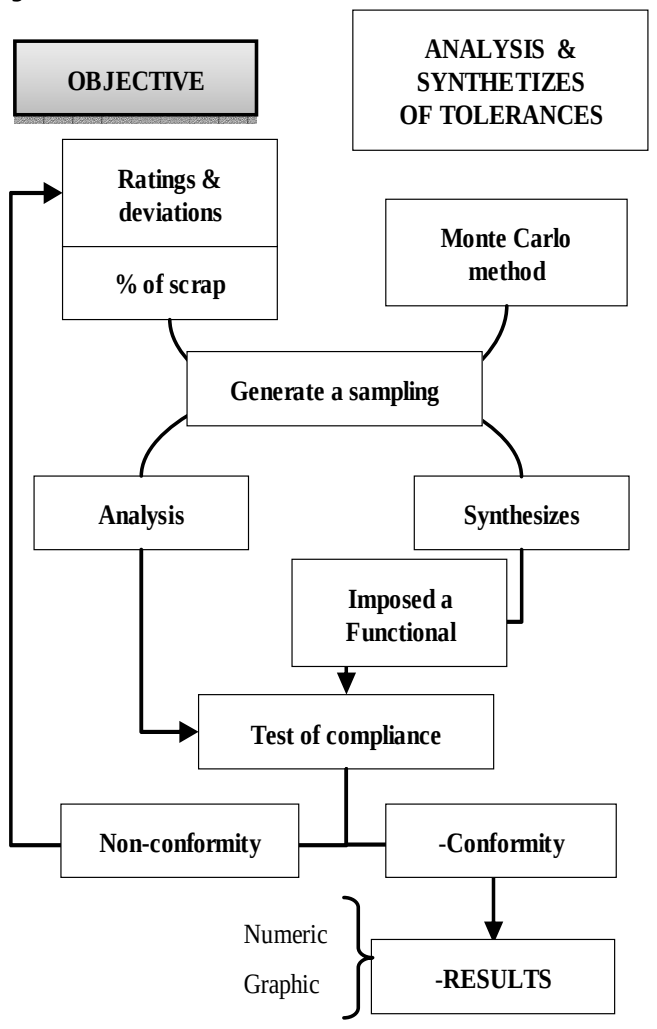

**Figure 7. Monte Carlo algorithm**

These parameters describe the various data made available for the manufacturer. In fact, the low probability of defective parts represents the number of parts when the costumer accepts. This % of scrap is at the origin of sampling test whose result depends on. It characterizes the step of optimization. Starting this instruction, the % of scrap imposed will be compared to that calculated;

- If given % of scrap < % of the scrap calculated : Expanding the strictest tolerances.

- If given % of scrap > % of scrap calculated:(Reduce tolerances widest.
- Otherwise, it retains the result.

The figure 7 describes an algorithm to apply a Monte Carlo method. This algorithm allows analysing and/or synthesising a chain of ratings. The identification of the approach and assumptions are completed by the party validation by examples. In the next paragraph, our objective is to identify the main feature of created approach. The validation will be done by a system of actuator clamping.

## V. VALIDATION

This part allows implementing the various features of the approach followed by an example to study chains ratings. It starts with an identification of data. This part is a modelling of chain of ratings (figure 8).

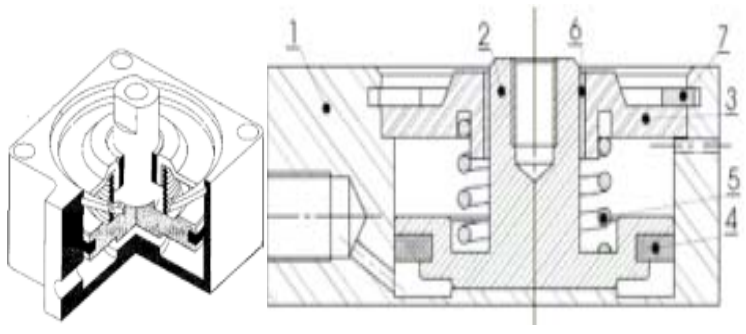

System of actuator clamping

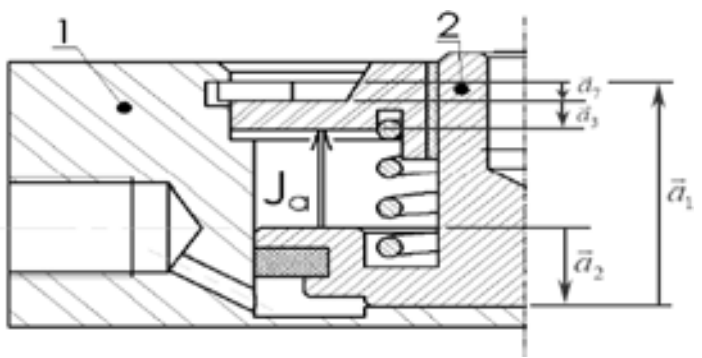

Ratings Chain of Assembly

$$a_1 = 25,3^{+0,5}_{-0} \quad ; a_2 = 9^{\pm 0,1}$$

$$a_3 = 4^{\pm 0,2} \quad ; a_7 = 2^{+0}_{-0,06}$$

**Figure 8 ; A system of actuator clamping**

For this example, we search to find a Functional rating (Ja) by the Worst Case method and the Monte Carlo method. Applying a first method, the result is;

$$J_{a\max} = a_{1\max} - (a_2 + a_3 + a_7)_{\min} = 11,16\ (mm)$$
$$J_{a\min} = a_{1\min} - (a_2 + a_3 + a_7)_{\max} = 10\ (mm)$$

$a_{i\max}, a_{i\min}$ ; maximal and minimal value of rating .

This result is limited to define the value of ratings in their extremes. So, this method can not provide information for the different combinations between ratings. It allows to identify the limit of ratings tolerances intervals.

| $IT_{a1}$ | $IT_{a2}$ | $IT_{a3}$ | $IT_{a7}$ | $IT_{Ja}$ |
|---|---|---|---|---|
| 0,5 | 0,2 | 0,4 | 0,06 | 1,16 |

By Monte Carlo method, the result is numerical. We obtain a table (figure 9-a) to represent a lot of production of parts. This table, modelled a number of assemblies, can be realised in manufacturing processes.

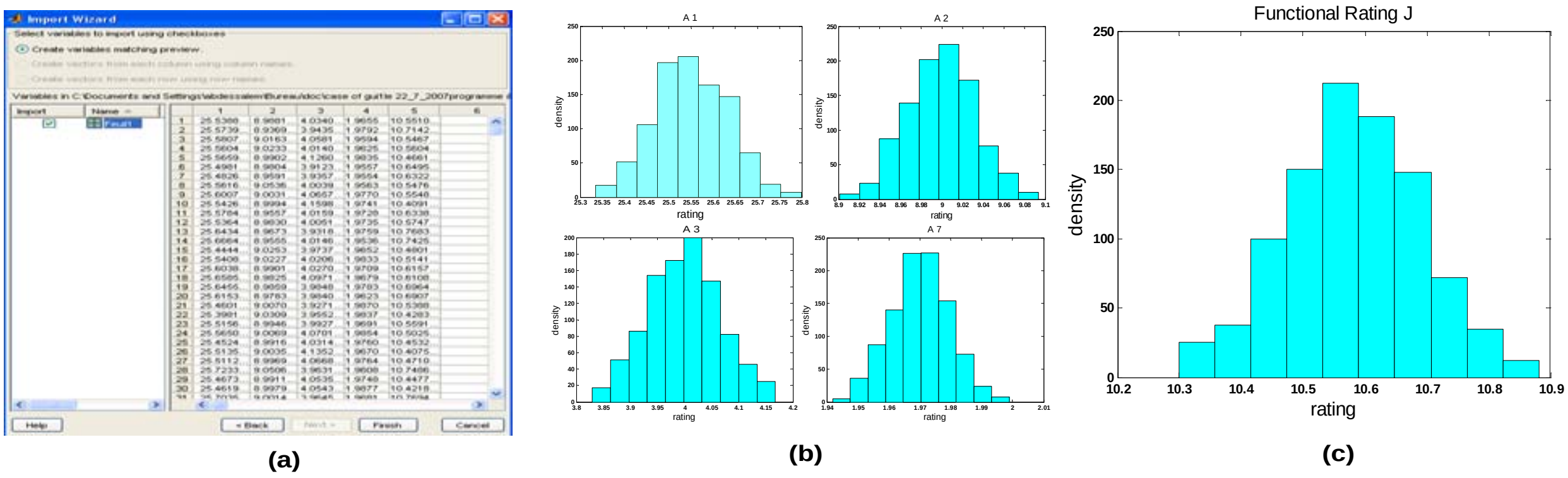

**Figure 9. a- Numeric result of ratings, b- Graphic result of ratings, c- Functional rating**

These results are to be represented by graphical figures (figure 9-b & 9-c). These figures present limits of variations of ratings. They provide a possibility of modification of value for some ratings. They can display the possible variation of ratings.
After this result, we can define a statistical interval of ratings:

| $IT_{a1}$ | $IT_{a2}$ | $IT_{a3}$ | $IT_{a7}$ |
|---|---|---|---|
| 0,0417 | 0,0067 | 0,0267 | 0,0006 |

And the statistical interval of Functional rating (Ja) is; $IT_{Ja} = 0{,}1628$.
Comparing the arithmetic intervals to statistical intervals, the statistical intervals are below then arithmetical intervals. $IT_{st} < IT_{ar}$. This result can provide to designer a possibility of enlarging the tolerances in the chain. It can give him possibility of performing a manufacturing process. So, it can reduce a number of defective parts. In fact, it can reduce a production costs by enlarging a critical tolerances [2]. This modification is preceded by a new simulation (sampling) according to a general algorithm (figure 7). We choose to modify a rating (a3). A new value is $a_3 = 4^{+0,15}_{-0,2}$. This value is accepted and can be found in a statistical interval of Functional rating equal ; $IT_{Ja} = 0{,}1595$. This procedure is repeated many times according to a % of scrap (figure 7).
In this step, they have identified the main features of our approach to specification and verification methods permitted by Worst Case method and Monte Carlo method which are complementary. This method allows the designer to specify the tolerances of chain of rating representing a mechanical assembly. The Monte Carlo method no gives the designer think in terms of statements limits of tolerance but a more realistic (production never gives state boundaries sides) and takes into account the context of production.

## VI. CONCLUSION

The development of new CAD systems that integrate effectively the various aspects needed to design a product (geometrical calculations, manufacturing ...), also requires taking into account the tolerancing aspect. In this paper, we presented an approach to help the designer in the specification and verification of tolerances. The worst case method (WCM) and the Monte Carlo method (MCM) are a complementary methods. The WCM is a method arithmetic based primarily on the total interchangeability of parts and assembly can feed the method of MCM by checking the configuration tolerances the worst case. Yet, the proposed approach is limited to tolerancing dimensional linear and to particular assumptions. It needs to introduce a deformability aspect of parts in assembly. This approach needs an ameliorate phase to integrate it in CAD / CAM environments.